\documentclass{revtex4}
\usepackage{amsfonts}
\usepackage{amsmath}
\usepackage{graphicx}
\usepackage{subfigure}

\begin{document}

\title{Vortex modes supported by spin-orbit coupling in a laser with
saturable absorption}
\author{Thawatchai Mayteevarunyoo$^{1}$, Boris A. Malomed $^{2,3}$, Dmitry
V. Skryabin$^{4}$ }
\affiliation{$^{1}$Department of Electrical and Computer Engineering, Faculty of
Engineering Naresuan University, Phitsanulok 65000, Thailand}
\affiliation{$^{2}$Department of Physical Electronics, School of Electrical Engineering,
Faculty of Engineering, and Center for Light-Matter Interaction, Tel Aviv
University, Tel Aviv 69978, Israel}
\affiliation{$^{3}$ITMO University, St. Petersburg 197101, Russia}
\affiliation{$^{4}$Department of Physics, University of Bath, Bath, BA2 7AY, UK}

\begin{abstract}
We introduce a system of two component two-dimensional (2D) complex
Ginzburg-Landau equations (CGLEs) with spin-orbit-coupling (SOC) describing
a wide-aperture microcavity laser with saturable gain and absorption. We
report families of two-component self-trapped dissipative laser solitons in
this system. The SOC terms are represented by the second-order differential
operators, which sets the difference, $|\Delta S|=2$, between the
vorticities of the two components. We have found stable solitons of two
types: vortex-antivortex (VAV) and semi-vortex (SV) bound states, featuring
vorticities $\left( -1,+1\right) $ and $\left( 0,2\right) $, respectively.
In previous works, 2D localized states of these types were found only in
models including a trapping potential, while we are dealing with the
self-trapping effect in the latteraly unconfined (free-space) model. The SV
states are stable in a narrow interval of values of the gain coefficients.
The stability interval is broader for VAV states, and it may be expanded by
making SOC stronger (although the system without SOC features a stability
interval too). We have found three branches of stationary solutions of both
VAV and SV types, two unstable and one stable. The latter one is an
attractor, as the unstable states spontaneously transform into the stable
one, while retaining vorticities of their components. Unlike previously
known 2D localized states, maintained by the combination of the trapping
potential and SOC, in the present system the VAV and SV complexes are stable
in the absence of diffusion. In contrast with the bright solitons in
conservative models, chemical potentials of the dissipative solitons
reported here are positive.
\end{abstract}
\maketitle

\section{Introduction}

The spin-orbit coupling (SOC), which was originally predicted in the form of
the Dresselhaus \cite{Dresselhaus} and Rashba \cite{Rashba1,Rashba2}
Hamiltonians, is a fundamentally important effect in physics of
semiconductors \cite{1,2,3}. More recently, much interest was drawn to the
possibility to emulate the SOC phenomenology, in its \textquotedblleft
clean" form, in spinor (two-component) atomic Bose-Einstein condensates
(BECs), by creating settings whose Hamiltonians can be mapped into the
Dresselhaus and Rashba forms (or a combination of both) \cite%
{Spielman1,Spielman3,Spielman2}, see also reviews \cite{dal,gal,zha}. In
these contexts, adequate mean-field SOC\ models are provided by systems of
nonlinear Gross-Pitaevskii equations (GPEs) coupled by linear-mixing terms
composed of first-order spatial derivatives.

Parallel to these developments, SOC effects were realized in the
exciton-polariton fields populating semiconductor microcavities operating in
the strong coupling regime \cite{Styopa1}-\cite{Styopa2}, and in many other
optical systems \cite{bli}, which are, however, less relevant in the present
context. It is important to note that SOC in polaritonic microcavities can
have two different physical origins, \textit{viz}., due to the underlying
SOC between excitons, or between the photonic modes. Thus, the existence of
polaritons is not required to observe SOC, and therefore the microcavities
operating in the regime of weak coupling between light and matter, when
polaritons are not formed, can be used to realize SOC effects. Below we are
make use of this opportunity, considering a planar microcavity that includes
ingredients providing both saturable gain and absorption \cite{tur,els,sel}.
This configuration is well known for its ability to maintain mode-locking of
transverse modes, leading to the formation of bright spatial solitons, as
well as to provide interplay of transverse and longitudinal modes leading to
3D effects, see, e.g., Refs. \cite{D1}-\cite{gen}. However, SOC effects have
not been so far purposefully studied in these settings. Ine microcavities,
the SOC originates from the splitting of resonances of the linearly
(in-plane) polarized modes, whose dominant electric-field components are
perpendicular (TE) and parallel (TM) to the in-plane component of the
carrier wave vector. Writing the field equations in terms of circular
polarizations produces the SOC terms with the second-order derivatives in
Eqs. (\ref{+}) and (\ref{-}), see below, which were previously discussed in
detail, see, e.g., \cite{fla}. Note that for transversely inhomogeneous
states (in particular, self-trapped solitons) the SOC terms are going to be
important even for states with the zero in-plane momentum.

A theoretical analysis of systems combining the cubic attractive
interactions of the BEC wave functions and linear spatial-derivative SOC in
two dimensions (2D) has revealed unexpected results. First, in the absence
of a trapping potential, SOC represented by the Rashba Hamiltonian produces
two species of \emph{stable} 2D solitons. These are \textit{semi-vortices}
(SVs, alias \textit{half-vortices} \cite{Pu}) with one zero-vorticity
component, and the other one carrying vorticity $S=\pm 1$, and \textit{mixed
modes} (MMs), so called because they mix zero-vorticity terms and ones with $%
S=\pm 1 $ in both components \cite{sak,Sherman1,Sherman2}. These results are
drastically different from the well-known ones obtained for traditional GPE
models in 2D, which produce \textit{Townes solitons} with zero vorticity
\cite{Townes}, and their ring-shaped vorticity-carrying extensions \cite%
{Yank}-\cite{skr}. The instability of the Townes solitons is driven by the
\textit{critical collapse }in the 2D space with the cubic attraction \cite%
{coll1,coll3}, while the vortex-ring solitons are subject to a still
stronger ring-splitting instability \cite{firth1,firth2}. The presence of
the SOC terms changes the situation, as they bring a coefficient which fixes
a length scale in the system, which, in turn, breaks the scaling invariance
that makes norms of the Townes' solitons degenerate (the entire family of
these solitons has a single value of the norm). The lifting of the norm
degeneracy pushes it below the degenerate value, which determines the
collapse-onset threshold, hence the collapse cannot occur anymore. This
mechanism secures the stabilization of both the SVs and MMs, and lends them
the role of the otherwise missing ground states \cite{Sherman1,Sherman2}.

A difference introduced by the microcavity SOC terms is their second-order
derivatives, giving rise to the vorticity difference between two spin
(circularly polarized) components to be $|\Delta S|=2$, rather than $|\Delta
S|=1$ as in the above-mentioned models of atomic BEC. In particular, the
analysis of the composite modes in the exciton-polariton system, which
includes a harmonic-oscillator trapping potential, was recently elaborated
in Ref. \cite{HS}. An essential result was the identification of stability
regions for MMs and vortex-antivortex (VAV) complexes, with vorticities $%
S=\pm 1$ in the two components (thus complying with the above-mentioned
constraint, $\Delta S=2$). An area of the MM-VAV bistability was identified
too, and stable SVs were found when the Zeeman splitting was added to the
two-component system.

While trapping potentials are an important ingredient in various systems,
see, e.g., \cite{bor1,sch} in addition to the above, it is also relevant to
construct stable composite modes supported by SOC in the laterally unbound
settings. In particular, laser systems with and without saturable absorbers
\cite{tur,els,gen,sel,jim} are known to support self-bound vortex rings \cite%
{gen,jim}. A possible practical realisation of a laser model considered
below is a wide-aperture semiconductor laser with a saturable absorber
section similar to the ones used in \cite{tur,els,sel}, where the SOC
effects originate from the above mentioned TE-TM splitting of the cavity
resonances. Theoretically, but without the SOC effects, the model used below
was elaborated in a series of papers by Rosanov and co-authors, see, e.g.,
\cite{saturable1,Vladimirov,rosprl}, and references there in. We generalise
the Rosanov's approach by including the SOC terms represented by second
derivatives mixing the two spin components. The system may also include
effects of diffusion, but, unlike the setting explored in Ref. \cite{HS} and
in many other models \cite{Crasovan}-\cite{we}, the presence of the
diffusion is not necessary for the stability of 2D states in the present
case. Below, we identify two species of stable self-bound vortex modes,
\textit{viz}., VAVs and SVs, both built of components obeying the constraint
of $\Delta S=2$. In particular, the VAVs feature two mutually symmetric
components with vorticities $S=\pm 1$, while the amplitude of the SV's
zero-vorticity component is much larger than the amplitude of its vortical
counterpart.

It is relevant to mention that VAVs are quite similar to two-component
states, produced by systems of 2D \cite{Nal,Styopa3,Raymond} and 3D \cite%
{Yaroslav} nonlinearily-coupled GPEs, with opposite vorticities, $\pm S$,
and identical density profiles in the components. In the latter context, the
bound states of this type were called ``hidden-vorticity modes", as, on the
contrary to their counterparts with equal vorticities in both components,
the total angular momentum of the hidden-vorticity states is zero. In
addition to that, conservative systems with the nonlinear
(cross-phase-modulation) interaction between two components (and no linear
coupling between them) maintain bound states similar to SVs, with zero and
nonzero vorticities in the components \cite{Technion1,Technion2}.

The subsequent presentation is organized as follows. The system of complex
Ginzburg-Landau equations (CGLEs) with the saturable gain, coupled by the
SOC terms, is introduced in Section II. The same section also reports some
simple analytical results (necessary conditions for the existence of stable
2D dissipative solitons). Numerical results, produced by the systematic
investigation of the 2D system, are reported in Section III, for two
above-mentioned species of stable modes, \textit{viz}., VAVs and MMs. The
paper is concluded by Section IV.

\section{The model and analytical estimates}

In terms of scaled time $t$ and coordinates $\left( x,y\right) $, the system
of CGLEs for wave functions $\psi _{\pm }$ of the two spin components, with
the linear loss, whose coefficient is scaled to be $1$, saturable gain and
saturable absorption, which are represented, respectively, by coefficients $%
g>0$ and $a>0$, and SOC linear mixing with real coefficient $\beta $, is
written as

\begin{eqnarray}
&&i\partial _{t}\psi _{+}=-(1-i\eta )(\partial _{x}^{2}+\partial
_{y}^{2})\psi _{+}+if\psi _{+}+\beta (\partial _{x}-i\partial _{y})^{2}\psi
_{-},  \label{+} \\
&&i\partial _{t}\psi _{-}=-(1-i\eta )(\partial _{x}^{2}+\partial
_{y}^{2})\psi _{-}+if\psi _{-}+\beta (\partial _{x}+i\partial _{y})^{2}\psi
_{+},  \label{-}
\end{eqnarray}
\begin{equation}
f=-1+\frac{g}{1+\varepsilon (|\psi _{+}|^{2}+|\psi _{-}|^{2})}-\frac{a}{
1+(|\psi _{+}|^{2}+|\psi _{-}|^{2})}.  \label{f}
\end{equation}
Here, the coefficient in front of the Laplacian, which represents the
diffraction, is also scaled to be $1$ (in this notation, physically relevant
values of the SOC coefficient, $|\beta |$, are definitely smaller than $1$),
and positive $\varepsilon <1$ defines a relative saturation strength of the
gain and absorption. The dispersive component of the linear loss (diffusion)
may be present, with coefficient $\eta >0$, but, in fact, the inclusion of
this term, whose physical origin may not be obvious, is not necessary for
producing stable 2D modes, therefore we set $\eta =0$ in what follows below.
The generic situation may be adequately represented by parameters
\begin{equation}
\varepsilon =0.1,~a=2,  \label{fixed}
\end{equation}
which implies weak saturation of the gain in comparison with absorption,
while the gain and SOC\ strengths, $g$ and $\beta $, will be varied as
physically relevant parameters controlling modes generated by the system.
Stationary states with real chemical potential $\mu $ are sought as
\begin{equation}
\psi _{\pm }=e^{-i\mu t}u_{\pm }\left( x,y\right) ,  \label{u}
\end{equation}
where complex stationary wave functions $u_{\pm }$ obey the following
equations:
\begin{gather}
\mu u_{+}=-(1-i\eta )(\partial _{x}^{2}+\partial _{y}^{2})u_{+}+iF\left(
x,y)\right) u_{+}+\beta (\partial _{x}-i\partial _{y})^{2}u_{-},  \label{u+}
\\
\mu u_{-}=-(1-i\eta )(\partial _{x}^{2}+\partial _{y}^{2})u_{-}+iF\left(
x,y\right) \psi _{-}+\beta (\partial _{x}+i\partial _{y})^{2}u_{+},
\label{u-}
\end{gather}
\begin{equation}
F\left( x,y\right) \equiv -1+\frac{g}{1+\varepsilon
(|u_{+}|^{2}+|u_{-}|^{2}) }-\frac{a}{1+(|u_{+}|^{2}+|u_{-}|^{2})}.  \label{F}
\end{equation}

For the analysis of stability of stationary states, we define perturbed
solutions as
\begin{equation}
\psi _{\pm }\left( x,y,t\right) =e^{-i\mu t}\left\{ u_{\pm }\left(
x,y\right) +\delta \left[ v_{\pm }\left( x,y\right) e^{\lambda t}+w_{\pm
}^{\ast }\left( x,y\right) e^{\lambda ^{\ast }t}\right] \right\} ,
\label{delta}
\end{equation}
where $\delta $ is a real infinitesimal amplitude of the perturbations, $%
\lambda $ is a complex eigenvalue, and complex perturbation eigenmodes, $%
v_{\pm }\left( x,y\right) $ and $w_{\pm }\left( x,y\right) $, satisfy the
linearized equations
\begin{gather}
\left( \mu +i\lambda \right) v_{+}=-\left( 1-i\eta \right) (\partial
_{x}^{2}+\partial _{y}^{2})v_{+}+\beta (\partial _{x}-i\partial
_{y})^{2}v_{-}+iF\left( x,y\right) v_{+}  \notag \\
+iu_{+}\left[ G\left( x,y\right) \left( u_{+}^{\ast
}v_{+}+u_{+}w_{+}+u_{-}^{\ast }v_{-}+u_{-}w_{-}\right) \right] ,
\label{BdG1}
\end{gather}
\begin{gather}
\left( \mu +i\lambda \right) v_{-}=-\left( 1-i\eta \right) (\partial
_{x}^{2}+\partial _{y}^{2})v_{-}+\beta (\partial _{x}+i\partial
_{y})^{2}v_{+}+iF\left( x,y\right) v_{-}  \notag \\
+iu_{-}\left[ G\left( x,y\right) \left( u_{+}^{\ast
}v_{+}+u_{+}w_{+}+u_{-}^{\ast }v_{-}+u_{-}w_{-}\right) \right] ,
\label{BdG2}
\end{gather}
\begin{gather}
\left( \mu -i\lambda \right) w_{+}=-\left( 1+i\eta \right) (\partial
_{x}^{2}+\partial _{y}^{2})w_{+}+\beta (\partial _{x}+i\partial
_{y})^{2}w_{-}-iF\left( x,y\right) w_{+}  \notag \\
-iu_{+}^{\ast }\left[ G\left( x,y\right) \left( u_{+}w_{+}+u_{+}^{\ast
}v_{+}+u_{-}w_{-}+u_{-}^{\ast }v_{-}\right) \right] ,  \label{BdG3}
\end{gather}
\begin{gather}
\left( \mu -i\lambda \right) w_{-}=-\left( 1+i\eta \right) (\partial
_{x}^{2}+\partial _{y}^{2})w_{-}+\beta (\partial _{x}-i\partial
_{y})^{2}w_{+}-iF\left( x,y\right) w_{-}  \notag \\
-iu_{-}^{\ast }\left[ G\left( x,y\right) \left( u_{+}w_{+}+u_{+}^{\ast
}v_{+}+u_{-}w_{-}+u_{-}^{\ast }v_{-}\right) \right] ,  \label{BdG4}
\end{gather}
where we have defined a real function,
\begin{equation}
G\left( x,y\right) \equiv \frac{a}{\left( 1+\left\vert u_{+}\right\vert
^{2}+\left\vert u_{-}\right\vert ^{2}\right) ^{2}}-\frac{\varepsilon g}{
\left( 1+\varepsilon \left( \left\vert u_{+}\right\vert ^{2}+\left\vert
u_{-}\right\vert ^{2}\right) \right) ^{2}}.  \label{G}
\end{equation}
As usual, the stability condition is $\mathrm{Re}(\lambda )\leq 0$ for all
eigenvalues.

The system of equations (\ref{BdG1})-(\ref{BdG4}) can be written as an
eigenvalue problem in the matrix form:
\begin{equation}
i\mathbf{L}\Psi =\lambda \Psi  \label{matrix}
\end{equation}%
where
\begin{equation}
\mathbf{L}=\left(
\begin{array}{cccc}
\mathbf{L}_{11} & \mathbf{L}_{12} & \mathbf{L}_{13} & \mathbf{L}_{14} \\
\mathbf{L}_{21} & \mathbf{L}_{22} & \mathbf{L}_{23} & \mathbf{L}_{24} \\
\mathbf{L}_{31} & \mathbf{L}_{32} & \mathbf{L}_{33} & \mathbf{L}_{34} \\
\mathbf{L}_{41} & \mathbf{L}_{42} & \mathbf{L}_{43} & \mathbf{L}_{44}%
\end{array}%
\right) ,\text{\ \ \ \ \ \ }\Psi =\left(
\begin{array}{c}
v_{+} \\
v_{-} \\
w_{+} \\
w_{-}%
\end{array}%
\right)  \label{L}
\end{equation}%
\begin{equation}
\mathbf{L}_{11,22}\equiv \left( 1-i\eta \right) (\partial _{x}^{2}+\partial
_{y}^{2})-iF\left( x,y\right) -iu_{+,-}\left[ G\left( x,y\right) \left(
u_{+/-}^{\ast }\right) \right] +\mu ,
\end{equation}%
\begin{equation}
\mathbf{L}_{12,21}\equiv -\beta (\partial _{x}-/+i\partial _{y})^{2}-iu_{+,-}%
\left[ G\left( x,y\right) \left( u_{-/+}^{\ast }\right) \right] ,
\end{equation}%
\begin{equation}
\mathbf{L}_{13,14}\equiv -iu_{+}\left[ G\left( x,y\right) \left(
u_{+,-}\right) \right] ,\mathbf{L}_{23,24}=-iu_{-}\left[ G\left( x,y\right)
\left( u_{+,-}\right) \right] ,
\end{equation}%
\begin{equation}
\mathbf{L}_{31,32}\equiv -iu_{+}^{\ast }\left[ G\left( x,y\right) \left(
u_{+,-}^{\ast }\right) \right] ,\mathbf{L}_{41,42}=-iu_{-}^{\ast }\left[
G\left( x,y\right) \left( u_{+,-}^{\ast }\right) \right] ,
\end{equation}%
\begin{equation}
\mathbf{L}_{33,44}\equiv -\left( 1+i\eta \right) (\partial _{x}^{2}+\partial
_{y}^{2})-iF\left( x,y\right) -iu_{+,-}^{\ast }\left[ G\left( x,y\right)
\left( u_{+/-}\right) \right] -\mu ,
\end{equation}%
\begin{equation}
\mathbf{L}_{34,43}\equiv \beta (\partial _{x}+,-i\partial
_{y})^{2}-iu_{+,-}^{\ast }\left[ G\left( x,y\right) \left( u_{-,+}\right) %
\right] .
\end{equation}%
The spectrum of the linear-stability operator $\mathbf{L}$ was constructed
by means of the Fourier collocation method.

To complete the formulation of the stability-analysis framework, we notice
that, being interested in stable dissipative solitons, a necessary condition
is the stability of the zero background in Eqs. (\ref{+}), (\ref{-}) and ( %
\ref{f}), which obviously amounts to condition $g<1+a$. On the other hand, a
necessary condition for the ability of the saturable gain to maintain
nontrivial modes is that the largest value of $f(n\equiv |\psi
_{+}|^{2}+|\psi _{-}|^{2})$ in Eq. (\ref{f}), which is attained at density
\begin{equation}
n_{0}=\frac{\sqrt{a}-\sqrt{\varepsilon g}}{\sqrt{\varepsilon }\left( \sqrt{g}%
-\sqrt{\varepsilon a}\right) },  \label{n0}
\end{equation}%
must be positive. The substitution of $n_{0}$ in Eq. (\ref{f}) yields a
lower bound for $g$, which, if combined with the above-mentioned upper one, $%
g<1+a$, defines the interval in which the gain coefficient may take its
values:
\begin{equation}
\sqrt{\varepsilon a}+\sqrt{1-\varepsilon }<g<1+a  \label{<g<}
\end{equation}%
[the compatibility condition for $a$ and $\varepsilon $, following from Eq.
( \ref{n0}), $\sqrt{\varepsilon a}+\sqrt{1-\varepsilon }<1+a$, always
holds]. An additional restriction on $g$ is imposed by the condition that
expression (\ref{n0}) must be positive too:
\begin{equation}
\varepsilon a<g<a/\varepsilon .  \label{gsecond}
\end{equation}%
Note that for values $a=2$ and $\varepsilon =0.1$ adopted here, interval ( %
\ref{<g<}) amounts to
\begin{equation}
1.396<g<3,  \label{1.4-3}
\end{equation}%
while interval (\ref{gsecond}) is much broader and may therefore be
disregarded.

Our objective is to constructs solutions of the CGLE system of Eqs. (\ref{+}
) and (\ref{-}) in the form of 2D bright solitons morphed as bound states of
two components with certain values of integer vorticities, $m-1$ and $m+1$,
so that they comply with the above-mentioned constraint, $\Delta S=2$. In
polar coordinates $\left( r,\theta \right) $, the relevant solutions with
real chemical potential $\mu $ can be defined as

\begin{equation}
\psi _{+}=\phi _{+}(r)\exp \left[ -i\mu t+i(m-1)\theta \right] ,~\psi
_{-}=\phi _{-}(r)\exp \left[ -i\mu t+i(m+1)\theta \right] ,  \label{vortex}
\end{equation}
with complex amplitude functions $\phi _{\pm }(r)$ satisfying the radial
equations:

\begin{gather}
\mu \phi _{+}=-(1-i\eta )\left[ \frac{d^{2}}{dr^{2}}+\frac{1}{r}\frac{d}{dr}%
- \frac{1}{r^{2}}\left( m-1\right) ^{2}\right] \phi _{+}  \notag \\
+if\phi _{-}+\beta \left( \frac{d^{2}}{dr^{2}}+\frac{2m+1}{r}\frac{d}{dr}+
\frac{m^{2}-1}{r^{2}}\right) \phi _{-},  \notag \\
\mu \phi _{-}=-(1-i\eta )\left[ \frac{d^{2}}{dr^{2}}+\frac{1}{r}\frac{d}{dr}%
- \frac{1}{r^{2}}\left( m+1\right) ^{2}\right] \phi _{-}  \notag \\
+if\phi _{-}+\beta \left( \frac{d^{2}}{dr^{2}}-\frac{2m-1}{r}\frac{d}{dr}+
\frac{m^{2}-1}{r^{2}}\right) \phi _{+},  \label{phiphi}
\end{gather}
where $f$ is defined by Eq. (\ref{f}), with $\left\vert \psi _{\pm
}\right\vert $ replaced by $\left\vert \phi _{\pm }\right\vert $.

The boundary condition for Eq. (\ref{phiphi}) at $r\rightarrow 0$ is that $%
\phi _{\pm }$ must be vanishing as $r^{\left\vert m\mp 1\right\vert }$ at $%
r\rightarrow 0$, except for the case of $m\mp 1=0$, when the boundary
condition is $d\phi _{\pm }/dr|_{r=0}=0$. At $r\rightarrow \infty $, soliton
solutions must feature the exponential decay,
\begin{equation}
\phi _{\pm }(r)\sim r^{-1/2}\exp \left( -\left( \lambda _{\mathrm{r}%
}+i\lambda _{\mathrm{i}}\right) r\right) ,  \label{infty}
\end{equation}%
with $\lambda _{\mathrm{r}}>0$ and, generally, a nonvanishing imaginary part
of the decay rate, $\lambda _{\mathrm{i}}$. The substitution of asymptotic
expression (\ref{infty}) in Eq. (\ref{phiphi}) and the linearization for the
exponentially small amplitude functions leads to the relation between the
imaginary and real parts,
\begin{equation}
\lambda _{\mathrm{i}}=-\frac{1+a-g}{2\left( 1\pm \beta \right) \lambda _{%
\mathrm{r}}},
\end{equation}%
and a quadratic equation for $\lambda _{\mathrm{r}}^{2}$,
\begin{equation}
4\left( 1\pm \beta \right) ^{2}\left( \lambda _{\mathrm{r}}^{2}\right)
^{2}+4\left( 1\pm \beta \right) \mu \left( \lambda _{\mathrm{r}}^{2}\right)
-\left( 1+a-g\right) ^{2}=0,  \label{lambda}
\end{equation}%
a larger root of which yields a relevant value, $\lambda _{\mathrm{r}}^{2}>0$
(here, sign $\pm $ is unrelated to the subscript of $\phi _{\pm }$). An
essential consequence of Eq. (\ref{lambda}) is that, while in the
conservative model, in which the linear-loss factor, $\left( 1+a-g\right) $,
does not appear, $\lambda _{\mathrm{r}}^{2}>0$ is only possible for a
negative chemical potential, $\mu <0$ (at least, in the physically relevant
case of $|\beta |<1$), in the dissipative system $\mu >0$ is admitted too.
Indeed, all the numerical solutions, reported below, have been obtained with
$\mu >0$. Further, it should be stressed that, as in the case of generic
dissipative solitons \cite{PS}-\cite{Kramer}, relevant solutions exist only
at isolated (positive) eigenvalues of $\mu $. The solitons are characterized
by the total integral power (norm),
\begin{equation}
N=\int \int \left[ \left\vert \psi _{+}\left( x,y\right) \right\vert
^{2}+\left\vert \psi _{-}\left( x,y\right) \right\vert ^{2}\right]
dxdy\equiv 2\pi \int_{0}^{\infty }\left[ \left\vert \phi _{+}\left( r\right)
\right\vert ^{2}+\left\vert \phi _{-}\left( r\right) \right\vert ^{2}\right]
rdr,  \label{Norm}
\end{equation}

The analysis is carried out below for two most essential species of the
soliton complexes, \textit{viz}., VAVs corresponding to $m=0$ (so called
because their components carry opposite vorticities, $-1$ and $+1$), and
SVs, corresponding to $m=1$, with component vorticities $0$ and $2$. For $%
m\geq 2$, bound states (\ref{vortex}) resemble \textquotedblleft excited
states" of SVs, introduced in the model of the atomic BEC in Ref. \cite{sak}
, and, as well as in that setting, it is plausible that they are completely
unstable.

\section{Numerical results}

Numerical solutions for the soliton modes were obtained by splitting complex
functions $\phi _{\pm }(r)\exp \left[ i(m\mp 1)\theta \right] $, defined in
Eq. (\ref{vortex}), into real and imaginary parts, and solving the resultant
equations in the Cartesian coordinates by means of the modified
squared-operator method \cite{Jianke}. The respective eigenvalues $\mu $
were found simultaneously with the stationary modes. Stability of the
stationary states was identified by means of systematic direct simulations
of perturbed evolution of the stationary states, governed by Eqs. (\ref{+})
and (\ref{-}), and, in parallel, by calculating the stability eigenvalues,
as determined by Eqs. (\ref{matrix}) and (\ref{L}).

\subsection{Vortex-antivortex (VAV) complexes, $m=0$}

\begin{figure}[tbp]
\centering\includegraphics[width=4in]{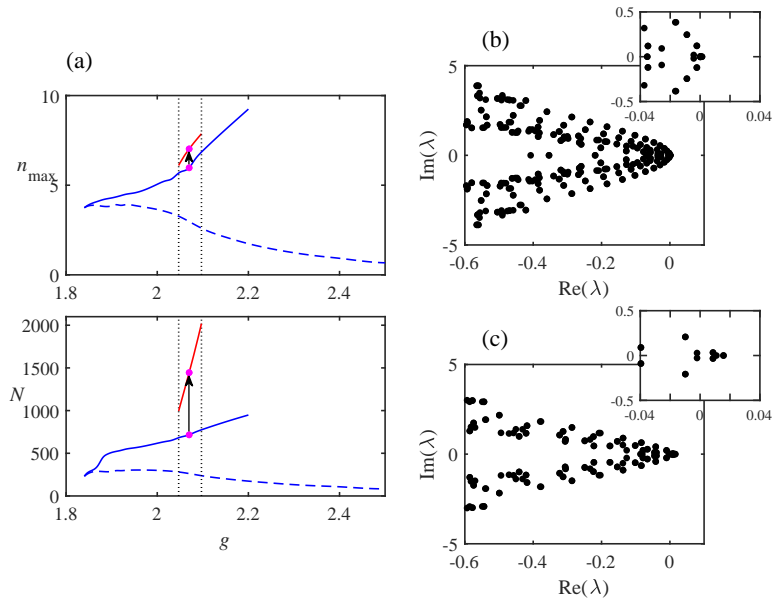}
\caption{(a) The peak density (top), defined as per Eq. (\protect\ref{nmax}%
), and the integral power (norm), defined as per Eq. (\protect\ref{Norm}),
of VAV (vortex-antivortex) modes vs. the gain coefficient, $g$, at the fixed
SOC coefficient, $\protect\beta =0.1$, with other parameters fixed as in Eq.
(\protect\ref{fixed}). The solid and dashed blue lines represent two
unstable families, and the red segment designates the stable one, in the
stability window (\protect\ref{stable}), marked by vertical dashed lines.
(b) The spectrum for the stable stationary VAV at $g=2.070$, which is marked
by the magenta dot belonging to the red branch in (a). (c) The spectrum for
the unstable VAV found at $g=2.070$, which is marked by the magenta dot
belonging to the blue branch in (a). In panel (b) and (c), insets zoom the
vicinity of $\protect\lambda =0$. }
\label{fig1}
\end{figure}

VAV modes, with $m=0$, were constructed as solutions of the stationary
equations, starting with the input adjusted to vorticities $\pm 1$ in the
two components,
\begin{equation}
\phi _{+}\left( r\right) =\phi _{-}(r)=\phi _{0}r\exp \left( -\alpha
r^{2}\right) ,  \label{input}
\end{equation}%
with some empirically chosen real parameters $\phi _{0}$ and $\alpha $, a
typical value being $\alpha =0.05$. As said above, parameters $\eta =0$, $%
a=2 $, $\varepsilon =0.1$, were fixed in Eqs. (\ref{+}), (\ref{-}) and (\ref%
{phiphi}), while the SOC and gain strengths $\beta $ and $g$ were varied. As
a result, three coexisting VAV\ families were found, two unstable and one
stable. Figure \ref{fig1} represents them by showing their integral power ( %
\ref{Norm}) and the peak density,
\begin{equation}
n_{\max }\equiv \max \left\{ \left\vert \psi _{+}\left( x,y\right)
\right\vert ^{2}+\left\vert \psi _{-}\left( x,y\right) \right\vert
^{2}\right\} ,  \label{nmax}
\end{equation}%
vs. $g$, at fixed $\beta =0.1$. In particular, the pair of unstable branches
(continuous and dashed blue lines in Fig. \ref{fig1}) emerge from a
bifurcation point at
\begin{equation}
g\approx 1.84.  \label{bif}
\end{equation}%
The stable (red) branch is shown only in the relatively narrow window where
it remains stable,
\begin{equation}
2.047<g<2.097,  \label{stable}
\end{equation}%
which is bounded by vertical dashed lines in Fig. \ref{fig1}. Note that,
although being much more narrow than interval (\ref{<g<}) determined by the
above-mentioned necessary conditions, stability region (\ref{stable}) is
located quite close to the center of the broad interval (\ref{<g<}). It is
relevant to note that, the narrow interval (\ref{stable}), and similar
intervals reported below, can be realized in terms of actual physical
parameters of laser cavities with the saturable gain and loss, as it follows
from Refs. \cite{jim,tur,els,gen}.

As concerns the chemical potential of the stable branch, in the stability
interval (\ref{stable}) it decreases nearly linearly, as a function of $g$,
from $\mu \left( g=2.047\right) =0.127$ to $\mu \left( g=2.097\right) =0.081$
. We stress that, as mentioned above, these values are positive, on the
contrary to necessarily negative chemical potential for solitons in
conservative models. As for the two unstable branches, they produce positive
and almost constant $\mu $ in the same interval (\ref{stable}).

\begin{figure}[tbp]
\centering\includegraphics[width=4in]{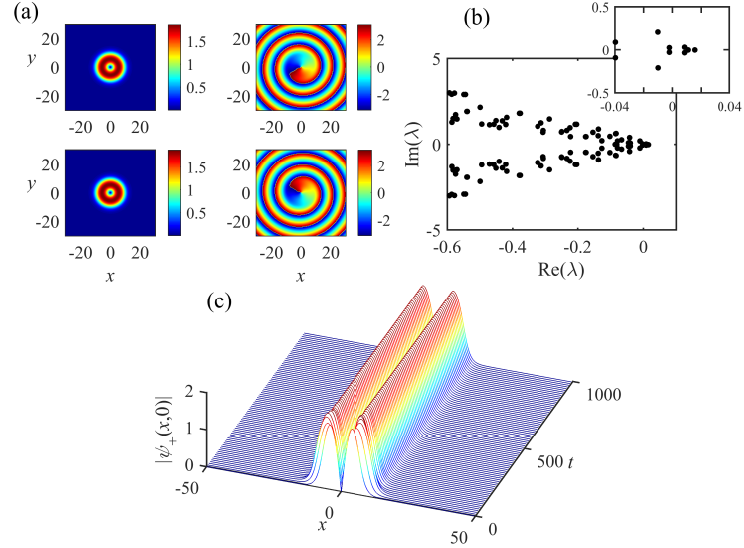}
\caption{The spontaneous transformation of an unstable VAV\ into its stable
counterpart at $g=2.07$, $\protect\beta =0.1$. (a) The top and bottom rows
displays the amplitude and phase structure of the $\protect\psi _{+}$ and $%
\protect\psi _{-}$ components, respectively, in the established stable VAV
complex. (b) Some eigenvalues in the stability spectrum of the initial
unstable VAV. The inset zooms the vicinity of $\protect\lambda =0$. (c) The
spontaneous transformation, by means of the cross section of one component, $%
\left\vert \protect\psi _{+}\left( x,0\right) \right\vert $. Both the
initial unstable and final stable VAVs are designated by dots in Fig.
\protect\ref{fig1}, with the spontaneous transition between them
schematically shown by the vertical arrows.}
\label{fig2}
\end{figure}

\begin{figure}[tbp]
\centering\includegraphics[width=4in]{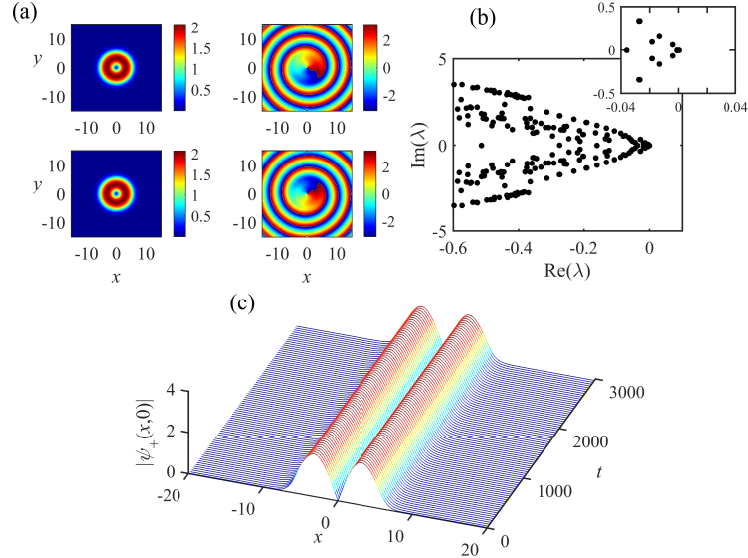}
\caption{(a) The top and bottom rows display the amplitude and phase
structure of components $\protect\psi _{+}$ and $\protect\psi _{-}$,
respectively, in a stable VAV mode found at $\protect\beta =0.8$ and $g=2.12$%
, with integral power $N=571.5$, peak density $n_{\max }=8.53$, and positive
chemical potential $\protect\mu =0.072$. (b) Some eigenvalues in the
stability spectrum of (a). The inset zooms the vicinity of $\protect\lambda %
=0$. (c) Stability of this mode in direct simulations of its perturbed
evolution.}
\label{fig3}
\end{figure}

\begin{figure}[tbp]
\centering\includegraphics[width=4in]{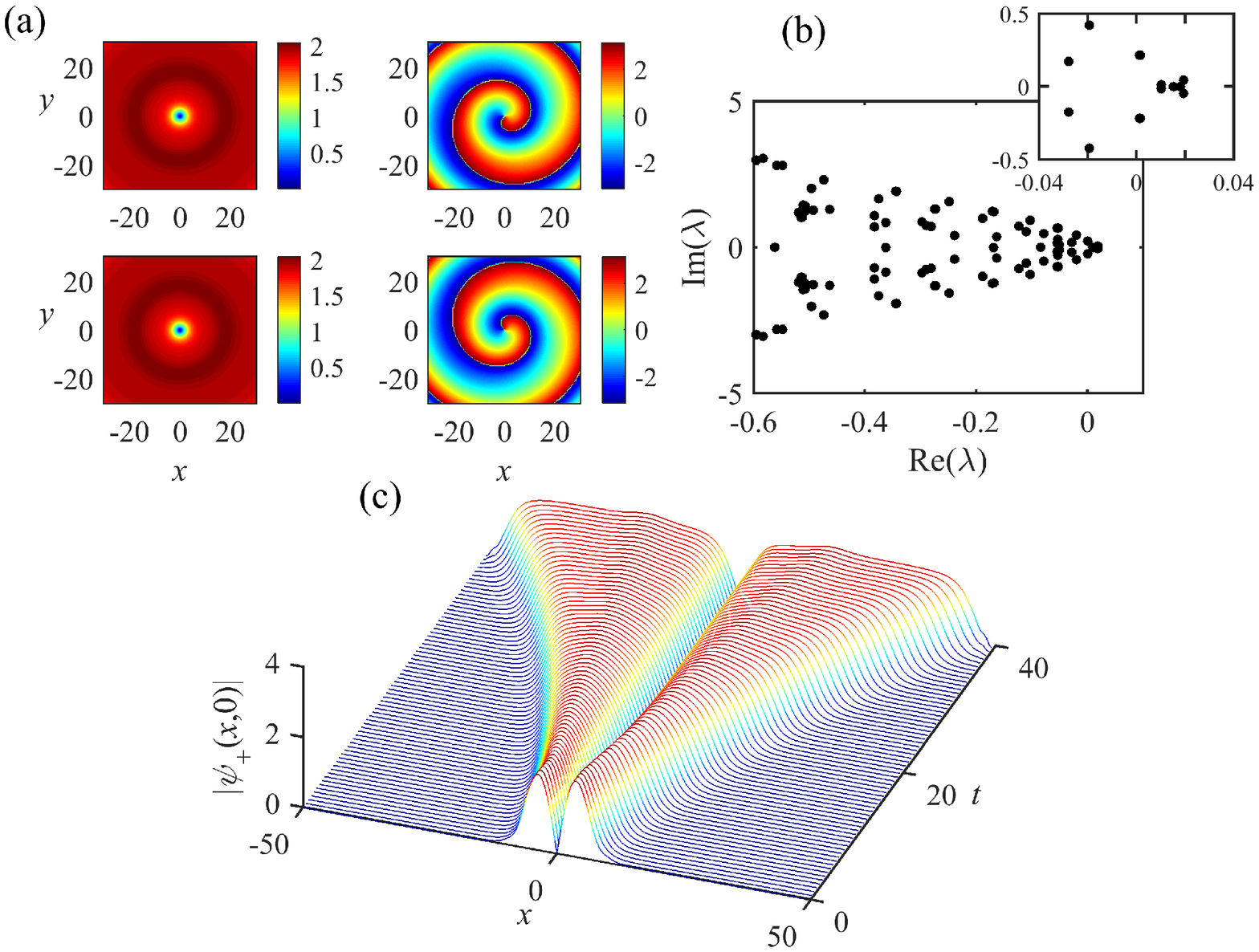}
\caption{The unstable VAV at $g=2.200$, $\protect\beta =0.1$. (a) The top
and bottom rows display the amplitude and phase structure of the $\protect%
\psi _{+}$ and $\protect\psi _{-}$ components, respectively. (b) Some
eigenvalues in the stability spectrum of the initial unstable VAV. The
insets zoom the vicinity of $\protect\lambda =0$. (c) The evolution of the
unstable VAV, which suffers blowup as the result of the instability
development. }
\label{fig4}
\end{figure}

At $g<2.047$, the development of the instability of the branches shown by
the blue lines in Fig. \ref{fig1} leads to their decay towards the zero
solution, while at $g>2.097$ the amplitude features exponential growth
(blowup, see Fig. \ref{fig4}). Inside stability interval (\ref{stable}), the
stable branch (the red segment in Fig. \ref{fig1}) is an \textit{attractor}:
solitons belonging to either unstable branch spontaneously transform into
ones belonging to the stable branch, as shown in Fig. \ref{fig2}. The
outcome of the transformation is identical to the VAV\ mode that may be
found as the stationary solution, see an example in Fig. \ref{fig3}. We did
not aim to extend the red branch in Fig. \ref{fig1} into the areas where it
is unstable, i.e., $g<2.047$ and $g>2.097$, see Eq. (\ref{stable}).

As concerns (in)stability eigenvalues, produced by the linearized equations
( \ref{matrix}) and (\ref{L}), typical examples of stable and unstable
spectra are displayed in Figs. \ref{fig2}(b), \ref{fig3}(b), \ref{fig4}(b)
and \ref{fig7}(b), respectively. These results are consistent with the
conclusions made on the basis of systematic direct simulations of the
perturbed evolution of the VAV modes.

The results of the investigation of the VAV modes are summarized in Fig. \ref%
{fig5}, which displays the stability region in the plane of the varying
parameters, $g$ and $\beta $ (the gain and SOC strengths). It is worthy to
note that the instability proceeds via the decay or blowup, without breaking
the axial symmetry of the VAVs. This is a drastic difference from the
above-mentioned hidden-vorticity modes, which are also built as bound states
of localized components with vorticities $S=\pm 1$ in the framework of the
conservative system of nonlinearly coupled GPEs, and are partly \cite%
{Nal,Styopa3,Raymond} or fully \cite{Yaroslav} unstable against the
splitting instability, which breaks the respective vortex ring into
fragments.
\begin{figure}[tbp]
\centering\includegraphics[width=3in]{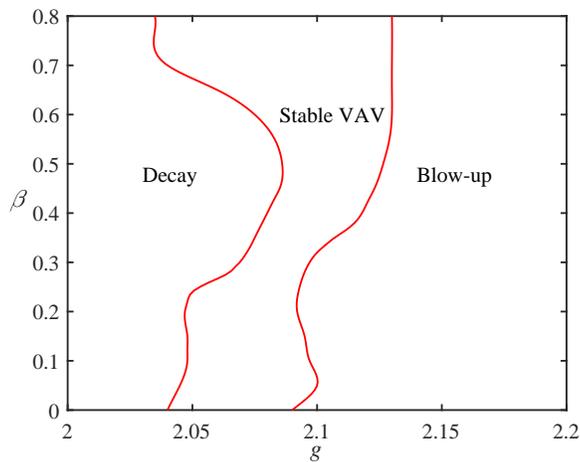}
\caption{The stability border for VAV modes in the plane of $\left( g,%
\protect\beta \right) $ for $\protect\eta =0$, $a=2.0$ and $\protect%
\varepsilon =0.1$.}
\label{fig5}
\end{figure}

Figure \ref{fig5} demonstrates that the stability window for the VAV modes
remains narrow in comparison with\ interval (\ref{1.4-3}) defined by the
above-mentioned coarse necessary conditions. The window exists too at $\beta
=0$:
\begin{equation}
2.040<g<2.090,  \label{stable1}
\end{equation}
when SOC is absent, Eqs. (\ref{+}) and (\ref{-}) being coupled only by the
saturable gain and loss, see Eq. (\ref{f}). With the increase of the SOC
coefficient up to relatively large values, $\beta \simeq 0.7$, the width of
the stability region expands by a factor $\simeq 2$, still staying quite
close to the center of the broad interval defined by necessary conditions ( %
\ref{1.4-3}). Systematically collected numerical data demonstrate that the
same conclusions remain true at other values of parameters $a$ and $%
\varepsilon $ in the underlying model based on Eqs. (\ref{+}) and (\ref{-}).

\subsection{Semi-vortex (SV) complexes, $m=1$}

Composite soliton modes corresponding to ansatz (\ref{vortex}) with $m=1$
were found, along with the corresponding eigenvalues $\mu $, as stationary
solutions initiated by input
\begin{gather}
\phi _{+}\left( x,y\right) =\phi _{0}\exp \left( -\alpha r^{2}\right) ,
\notag \\
\phi _{-}\left( x,y\right) =\left( \phi _{0}/10\right) r^{2}\exp \left(
-\alpha r^{2}\right) ,  \label{Init_cond1}
\end{gather}%
Unlike input (\ref{input}) which generates the VAV modes, here the amplitude
of the vortex component is taken essentially smaller than in the
zero-vorticity component, because the vortex component in established SV
modes tends to have a relatively small amplitude \cite{sak}-\cite{Sherman2}.

\begin{figure}[tbp]
\centering\includegraphics[width=3in]{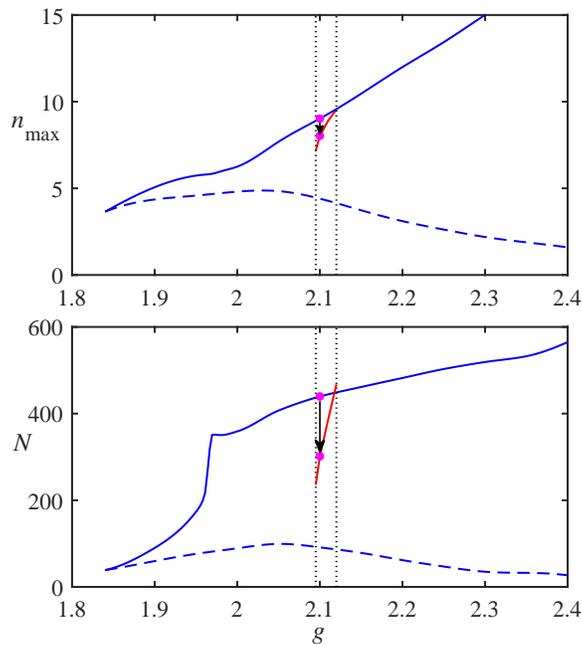}
\caption{The same as in Fig. \protect\ref{fig1}, but for SV families, with
vorticities $0$ and $2$ of the two components. The parameters in Eqs. (
\protect\ref{+}) and (\protect\ref{-}) are the same as in Fig. \protect\ref%
{fig1}, i.e, $\protect\eta =0$, $\protect\varepsilon =0.1$, $a=2$, and $%
\protect\beta =0.1$. The narrow stability window (\protect\ref{stable2}) is
exhibited by this figure. The dots connected by the vertical arrow
correspond to the spontaneous transformation of an unstable SV into a stable
one, as shown in Fig. \protect\ref{fig7}. }
\label{fig6}
\end{figure}

The systematic numerical analysis yields a stability band for the SV modes
shown in Fig. \ref{fig6}, which demonstrates a situation generally similar
to that displayed for VAV modes in Fig. \ref{fig1}, but with a stability
window,
\begin{equation}
2.095<g<2.120,  \label{stable2}
\end{equation}
whose width is half of that defined by Eq. (\ref{stable}) for the VAVs. This
interval, similar to its counterpart (\ref{stable}), stays close to the
center of the broad interval (\ref{1.4-3}), which is defined by the
necessary conditions considered above.

In the present case too, three branches of stationary solutions are observed
in Fig. \ref{fig6}, \textit{viz}., two unstable ones, which emerge at the
bifurcation point, that virtually exactly coincides with its counterpart for
the VAV states, given by Eq. (\ref{bif}), and a stable branch shown by the
red segment in Fig. \ref{fig6}. Another similarity to the case of VAV modes
is that, outside the stability band (\ref{stable2}), unstable SVs decay to
zero at $g<2.095$, and undergo blowup at $g>2.120$. On the other hand, the
comparison of Figs. \ref{fig1} and \ref{fig6} shows that the integral power
of the stable SVs is smaller than the power of the stable VAVs, roughly, by
a factor of $5$.

The chemical potential of the stable branch decreases nearly linearly, as a
function of $g$, in the stability interval (\ref{stable}), from $\mu \left(
g=2.095\right) =0.159$ to $\mu \left( g=2.120\right) =0.106$, while $\mu $
remains nearly constant in interval (\ref{stable2}) for both unstable
branches. Note that all these values of the chemical potential are positive,
as they are for the VAV branches considered above.

\begin{figure}[tbp]
\centering\includegraphics[trim=0 0.85in 0 0 ,clip,width=4.5in]{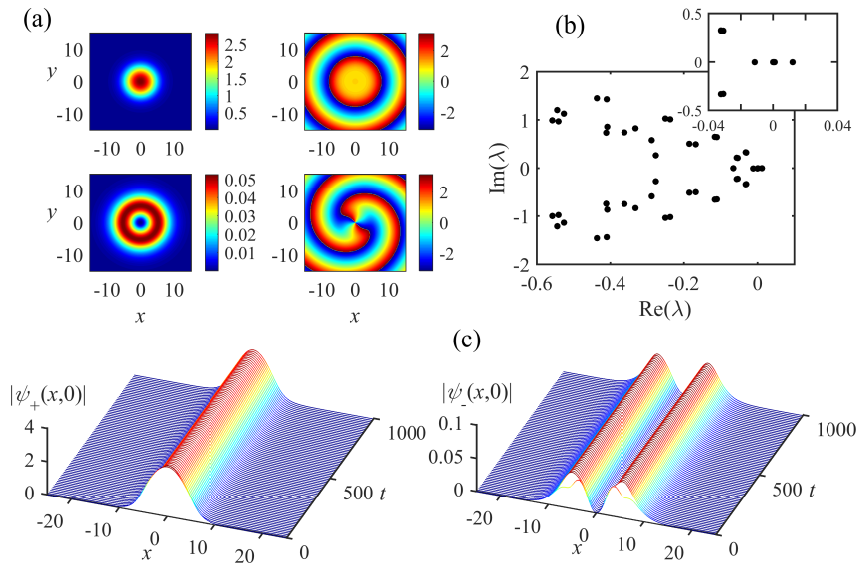}
\caption{(a) The top and bottom rows display the amplitude and phase
structures of the zero-vorticity and vortical components ($\protect\psi _{+}$
and $\protect\psi _{-}$, respectively), for a stable SV mode found at $%
\protect\beta =0.1$ and $g=2.10$, with other parameters fixed as per Eq. (%
\protect\ref{fixed}). (b) Some eigenvalues in the stability spectrum of the
initial unstable SV. The insets zoom the vicinity of $\protect\lambda =0$.
(c) The evolution of $\left\vert \protect\psi _{+}\left( x,0\right)
\right\vert $ and $\left\vert \protect\psi _{-}\left( x,0\right) \right\vert
$, which displays, at the same parameters, the spontaneous transformation of
an unstable SV into the stable one. Both the initial unstable and final
stable SVs are designated by dots in Fig. \protect\ref{fig6}, with the
spontaneous transition between them designated by the vertical arrows.}
\label{fig7}
\end{figure}

As well as in the case of VAV modes, the stable branch is, within the limits
imposed by Eq. (\ref{stable2}), an attractor, as initial states
corresponding to either of the two unstable families spontaneously transform
into the stable counterpart, keeping the SV structure. A typical example of
the spontaneous transformation is displayed in Fig. \ref{fig7}.

\subsection{The system without the SOC terms}

At $\beta =0$, the vortex component of the SV vanishes, $\psi _{-}=0$, while
the zero-vorticity one turns into am axisymmetric dissipative soliton
produced by the single equation (\ref{+}):
\begin{equation}
i\partial _{t}\psi _{+}=-(1-i\eta )(\partial _{x}^{2}+\partial _{y}^{2})\psi
_{+}+i\left( -1+\frac{g}{1+\varepsilon |\psi _{+}|^{2}}-\frac{a}{1+|\psi
_{+}|^{2}}\right) \psi _{+}  \label{single}
\end{equation}%
(recall we actually set $\eta =0$). In the case of $\beta =0$, essentially
the same equation, with $\psi _{+}$ replaced by $\tilde{\psi}_{+}$, applies
to the two-component system with equal components, of substitution $\psi
_{\pm }=(1/\sqrt{2})\tilde{\psi}_{+}$. The numerical solution of Eq. (\ref%
{single}) produces the respective stability window,
\begin{equation}
2.090<g<2.130,  \label{stable3}
\end{equation}%
which is somewhat broader than its counterpart (\ref{stable2}), found for $%
\beta =0.1$.

\begin{figure}[tbp]
\centering\includegraphics[width=4in]{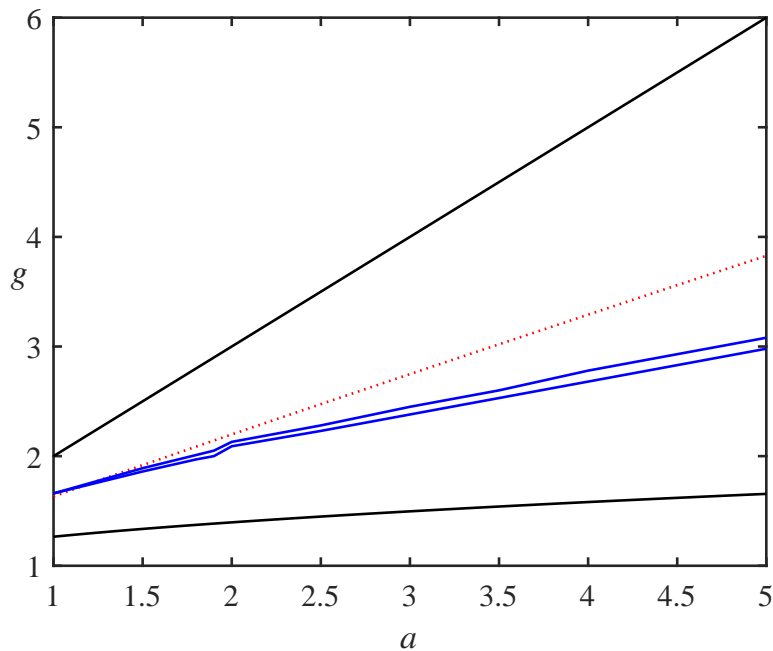}
\caption{The stability diagram in the plane $(a,g)$ with $\protect%
\varepsilon =0.1$. The black solid lines are plotted by Eq. (\protect\ref%
{<g<}) while the red dotted line is the middle of the interval. The
stability region for the zero vorticity soliton corresponds to the space
between the blue lines. The region of the stability vanish at $a=1$.}
\label{Fig8}
\end{figure}

\begin{figure}[tbp]
\centering\includegraphics[width=4in]{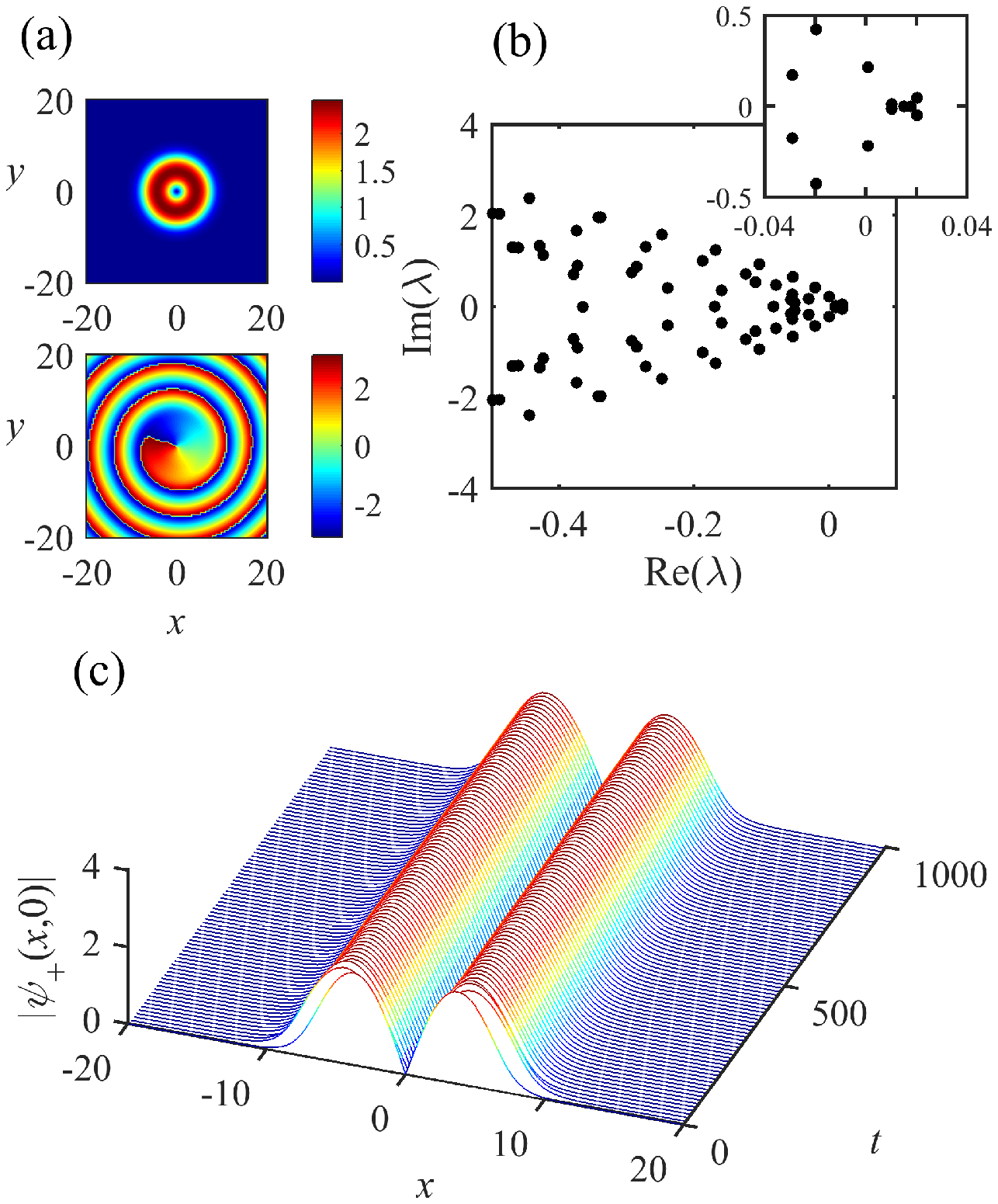}
\caption{An example of the zero vorticity soliton for $g=2.75$ and $a=4.0$.
(a) The amplitude and phase of $\protect\psi _{+}$ component (at $t=1000$).
(b) Some eigenvalues in the stability spectrum of the initial unstable zero
vorticity. The insets zoom the vicinity of $\protect\lambda =0$. (c) The
evolution of $\left\vert \protect\psi _{+}\left( x,0\right) \right\vert $
showing the spontaneous transformation of unstable zero voriticity soliton
into a stable one.}
\label{Fig9}
\end{figure}

The relative narrowness of the stability windows (\ref{stable2}) and (\ref%
{stable3}) in comparison with the ones given by Eqs. (\ref{stable}) and (\ref%
{stable1}) for VAVs suggests that SV modes are more fragile in comparison
with the VAVs. This expectation is confirmed by the fact that, on the
contrary to the situation for the VAVs, whose stability area tends to expand
with the growth of the SOC strength $\beta $ in Fig. \ref{fig3}, the
increase of $\beta $ leads to shrinkage of the SV stability window, which
closes and does not exist at $\beta \geq 0.24$ (not shown here in detail).
Finally, Fig. \ref{Fig8} summarizes the findings in the plane of $\left(
a,g\right) $ for fixed $\varepsilon =0.1$. It is clearly seen that the
region of stable zero vorticity soliton is relatively broad at large values
of $a$. Typical exaple of the spontaneous transformation of unstable zero
vorticity soliton is displayed in Fig. \ref{Fig9}.


\section{Conclusion}

The objective of this work is to construct 2D self-trapped states
(dissipative solitons) in the system of CGLEs with saturable gain and
absorption for SOC (spin-orbit-coupled) modes of an optical microcavity with
saturable gain and loss, in the absence of any geometric trapping. The
second-order linear differential operator representing SOC\ creates
complexes with difference $\Delta S=2$ between vorticities of the two
components. Previously, 2D localized dissipative vortex states were
predicted, under the action of SOC, in the presence of a trapping potential,
but they were not found in the free space. Here, we have identified stable
complexes of the VAV (vortex-antivortex) and SV (semi-vortex) types, i.e.,
ones with vorticities $\left( -1,+1\right) $ and $\left( 0,2\right) $ in the
two components. The 2D solitons of the latter type are quite fragile, being
stable in a very narrow (but, nevertheless, existing) window. For the VAVs,
the stability interval, defined in terms of the gain coefficient, is rather
narrow too, but it may be expanded by applying stronger SOC.

It is worthy to note that, on the contrary to the previously reported 2D
localized states of the VAV, MM (mixed-mode), and SV types, supported by the
trapping potential \cite{HS}, as well as single-component trapped \cite{we}
and self-trapped \cite{Crasovan,Herve}\ vortices, the stability of the VAVs
and SVs in the present system (including its simplified form which does not
include SOC) does not require the presence of diffusion terms (dispersive
linear losses, whose physical origin may be problematic) in the CGLE model.
Due to the absence of the diffusion, the 2D dissipative system introduced in
this work keeps the Galilean invariance, which suggests a possibility to
introduce moving solitons and simulate collisions between them (cf. Ref.
\cite{Fukuoka}), as well as a possibility of forming bound states of two or
several dissipative solitons. These issues should be a subject of a separate
work.

\section*{Funding}

Thailand Research Fund (grant BRG6080017); The Royal Society (grant IE
160465); Israel Science Foundation (grant 1286/17); ITMO University Visiting
Professorship (Government of Russia Grant 074-U01).

\end{document}